# Mechanisms behind slow photo-response character of Pulsed Electron Deposited ZnO thin films

Mehmet Özdoğan[1,2], Gökhan Utlu[1], and Cem Çelebi[2*]

[1]Department of Physics, Faculty of Science, Ege University, 35100, Izmir, Turkey

[2]Quantum Device Laboratory, Department of Physics, Izmir Institute of Technology, 35430, Izmir, Turkey

[*]e-mail address: cemcelebi@iyte.edu.tr

**ABSTRACT:** Zinc Oxide (ZnO) semiconductor is ideal candidates for ultra-violet (UV) photodetector due to its promising optoelectronic properties. Photodetectors based on ZnO nanostructures show very high photoconductivity under UV light, but they are plagued by slow photo-response time as slow as several tens of hours, even more. Most of the studies claimed that atmospheric adsorbates such as water and oxygen create charge traps states on the surface and remarkably increase both the photoconductivity and response time, but there are also limited studies that claiming the defect induced states acting as hole trap centers responsible for these problems. However, the underlying physical mechanism is still unclear. Here we study the effects of both adsorbates and defect-related states on the photo-response character of Pulsed Electron Deposited ZnO thin films. In order to distinguish between these two mechanisms, we have compared the time-dependent photo-response measurements of bare-ZnO and $SiO_2$ encapsulated-ZnO thin film samples taken under UV light and high vacuum. We show that the dominant mechanism of photo-response in ZnO is the adsorption/desorption of oxygen and water molecules even when the measurement is performed in high vacuum. When the samples are encapsulated by a thin $SiO_2$ layer, the adsorption/desorption rates can significantly improve, and the effects of these molecules partially removed.

*Keywords: Zinc Oxide, Pulsed Electron Deposition, photo-response, $SiO_2$, encapsulation, defects, atmospheric adsorbates.*

## 1. Introduction

Zinc Oxide (ZnO) material has received considerable attention due to its direct wide bandgap (3.37 eV) and high exciton binding energy (60 meV) for optoelectronic applications such as ultra-violet (UV) light detection for a long time [1]. Numerous studies have been conducted to achieve realization of electrically stable nano-electronic and nano-optoelectronic devices based on ZnO that are expected to operate a long time with reduced noise levels under atmospheric conditions; however, the electrical and optoelectronic properties of ZnO material are often changed over time [2,3]. Nanostructured or thin film forms of ZnO material are highly sensitive to environment because of their surface defects serving adsorption sites for atmospheric gases on the surface [4], which can be utilized sensitive gas detection performance [5]. Reactive gases like $O_2$ and $H_2O$ in ambient can act as electron trap centers when they are stuck on the surface of n-type materials such as ZnO. Unintentional adsorption/desorption of these molecules can randomly modify the electronic transport characteristics of these kinds of materials [6–9]. The electrical conductivity of n-type ZnO material decreases upon the adsorption of these gas molecules since the main charge carriers (electrons) are captured, or increases upon desorption of these molecules from the surface [10,11]. It has been shown that the adsorbates like $O_2$ and $H_2O$ can significantly change the carrier density of ZnO and strongly dominate the photo-response characteristics of UV sensitive photodetectors fabricated out of ZnO thin films or other forms [12]. Therefore, it is of great importance to unveil the detailed physical mechanism underlying adsorbate induced electrical changes prior to designing solid-state devices out of ZnO n-type semiconductor.

In ZnO materials, high photoconductivity and slow photo-responsivity characters to the UV light have been simultaneously observed not only due to their high sensitivity to atmosphere [13] but also their intrinsic defects acting as hole trap centers within the bandgap that prolong their response time [14]. Most of the studies suggested that the slow photo-response character of ZnO originates from atmospheric adsorbates such as oxygen and water molecules that create charge trap states on the surface. For example, Ahn *et al.* [15] proposed that the slow photo-response in sol-gel synthesized ZnO nanowires is due to the charge transfer between water molecules and ZnO nanowire. Also, Soci *et al.* [6] obtained very high photoconductive gain as high as $G \sim 10^8$ for ZnO nanowire by means of desorption of oxygen. Moreover, Li *et al.* [10] proposed that the reason is due to the both effects of oxygen and water molecules. On the other hand, there are also limited studies that claiming the suspects are defective states behaving hole trap centers that prolong the response time. For instance, Moazzami *et al.* [14] reported that the photo-response of ZnO epilayers was dominated by shallow and deep trap states. However, the reason of higher but slower photo-response character of ZnO cannot be explained via only adsorption/desorption of atmospheric species ($O_2$ and/or $H_2O$) or defect-induced levels within the bandgap. It should be due to combination of both effects. In previous work [16], we have shown that in high vacuum, the photo-response was bigger and slower than that in air, confirming the atmospheric adsorbates such as $O_2$ and $H_2O$ greatly dominate the photo-responsivity. And, they suppress the effects of defect states. In this work, therefore, we have developed an experimental approach to distinguish the effect of defect-states acting as hole trap centers from the effect of atmospheric adsorbates by simply encapsulation of ZnO thin film surface with a thin $SiO_2$ layer. We show that encapsulation of ZnO thin films with $SiO_2$ partially removes the contribution of adsorbates to photo-response with a substantial reduction of adsorption and desorption times.

The objective of this work is to explain the aforementioned mechanisms more comprehensively and systematically, thus this study contributes to a better understanding of the ambient and intrinsic defects effects on the slow photo-response character of ZnO nanostructures than previously reported works, and helps guide the design of ZnO-based photodetectors or gas sensors. Additionally, many encapsulation approaches have been employed for 2D materials such as $HfO_2$ [17] and PMMA [18] for $MoS_2$, $SiO_2$ for graphene [19]; yet, there is no detailed study investigating encapsulation effect on the photo-response properties of ZnO thin film.

## 2. Experimental Details

Before the deposition of ZnO thin films, Cr (3 nm)/Au (80 nm) source/drain electrodes were patterned on 10 mm x 10 mm fused-quartz substrates by thermal evaporation technique, and the inset of Fig. 1(b) shows a typical device schematic. The channel length between source and drain electrodes was set to 200 µm, and this configuration was used to Transient Photocurrent Spectroscopy (TPS) measurements. Employing a shadow mask, 185 nm thick ZnO films with a size of 4 mm x 4 mm were deposited at the center of the substrates by Pulsed Electron Deposition (PED) method. The details of PED technique can be found in Ref. [20]. The deposition parameters were fine-tuned to achieve the best ablation and plume (i.e., plasma and evaporated material) intensity, and to get high crystal quality of ZnO films. The deposition of ZnO thin films were done with a substrate temperature of 400 °C, an oxygen pressure of 1.57 Pa, an electron discharge voltage of 15 kV, and a pulse frequency of 5 Hz. Then, on the top of ZnO thin film, a thin layer of $SiO_2$ (~200 nm thick and 5 mm x 5 mm sized) was deposited using PED system as an encapsulation layer to assure full coverage of surface. For comparison, a set of bare ZnO thin films were also deposited by using the same protocols mentioned above.

After the successful fabrication of ZnO thin films, the structural, electrical, and opto-electronic properties of samples have been evaluated by different characterization techniques. Scanning Electron Microscopy (SEM) was performed using Quanta 250 instrument with typically operated at 5.0 kV to characterize the ZnO thin film morphologies.

The XPS analyses were performed using a Thermo Scientific Model K-Alpha XPS instrument with monochromatic Al Kα radiation (1486.68 eV) under operating pressure set to $2 \times 10^{-7}$ Pa. The survey spectrum scan was completed by taking the average of 10 scans with X-ray spot size at 300 μm and passing energy at 30 eV. All the XPS spectra were calibrated by taking the C 1 s peak located at ~284 eV as a reference. Data were analyzed using Avantage XPS software package. Peak fitting was done using Shirley/Smart type background and Gaussian/Lorentzian convolution shapes.

The X-ray diffraction (XRD) was carried out via a Philips X'Pert Pro Theta/2Theta Diffractometer with a copper K-Alpha X-ray source (λ=1.540 Å). The scans were performed from 20° to 80° in 0.001° steps.

The Photoluminescence (PL) spectra of the samples were taken by a Perkin Elmer LS-55 Luminescence Spectrometer (with a pulsed Xenon lamp) at an excitation wavelength of 350 nm with both 10 nm excitation and emission slit widths at room temperature. The excitation energy of 3.57 eV (350 nm) was slightly above the ~3.26 eV typical bandgap of our PED-fabricated samples.

Electrical and opto-electronic characterizations of the samples were done under 254 nm wavelength UV light illumination (output power of 3 mW) inside a high vacuum chamber with a base pressure of about $4 \times 10^{-3}$ Pa. For TPS experiments, an electronic shutter mechanism is coupled to the UV light source. The photocurrent data of the samples were acquired by using Keithley 6485 Picoammeter, and Keithley 2400 Source Meter. During measurement, the applied voltage between source and drain was kept constant at 0.5 V. Prior to the each set of measurements, we conducted I-V measurements for bare and encapsulated ZnO thin films (denoted as B-ZnO and E-ZnO, respectively) before and after TPS measurements, and then the samples were left overnight under ambient conditions in order to return back to their initial states. Afterwards, the samples were exposed to UV light for about 3 h. in high vacuum to remove existing adsorbates from the surface. Right after 3 h of continuous UV light exposure, TPS measurements were conducted for short periods as 30 s with three on/off cycles to reveal time dependent photo-response characteristics of the samples.

## 3. Results

The SEM image of ZnO thin film exhibits nanoparticles appearing as white spots on the film surface as seen in Fig. 1(a). These particles are more likely to occur as a result of thin film fabrication with high energy deposition techniques such as Pulsed Laser Deposition (PLD) and PED [21,22]. However, the origin of such nanoparticles is a matter of discussion. They can be directly emitted from target during ablation process, or they can be occurred in the gas phase, during transfer of species from the target toward the substrate [23]. In addition, we conducted EDX analysis to identify the elemental compositions of these nanoparticles, and the results showed that they possess almost same compositions of other regions on the film surface. The presence of such particles on the film surface increases the surface area-to-volume ratio, which maximizes the interaction between surface and atmospheric gases, and makes it more sensitive to the environment.

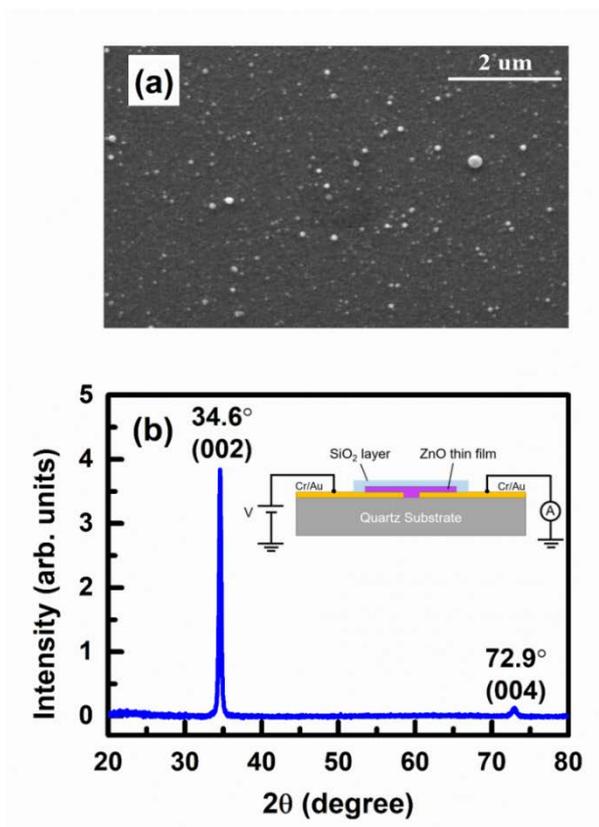

**Fig. 1. (a)** SEM image and **(b)** XRD pattern of bare ZnO thin film. The inset of **(b)** shows typical device schematic of SiO$_2$ encapsulated ZnO device.

Fig. 1(b) displays the XRD pattern of B-ZnO thin film. As seen in XRD graph, the PED-fabricated ZnO thin films have a single sharp peak at an angle of around 34.6° and very small peak around 72.9° which were identified as (002) and (004) orientations of hexagonal wurtzite ZnO crystal structure. It indicates that the ZnO thin film was deposited along a c-axis orientation of the quartz substrate. This c-axis orientation of crystallites is classically observed for ZnO films, whatever the method of deposition and the substrate nature [20]. The average crystallite size (D) of ZnO thin films was calculated using the (002) major diffraction peak by Debye-Sherrer formula ($D = 0.89\lambda/\beta cos\theta$), and it was found to be 30 nm. This value is comparable to that of PLD-fabricated ZnO thin films [24].

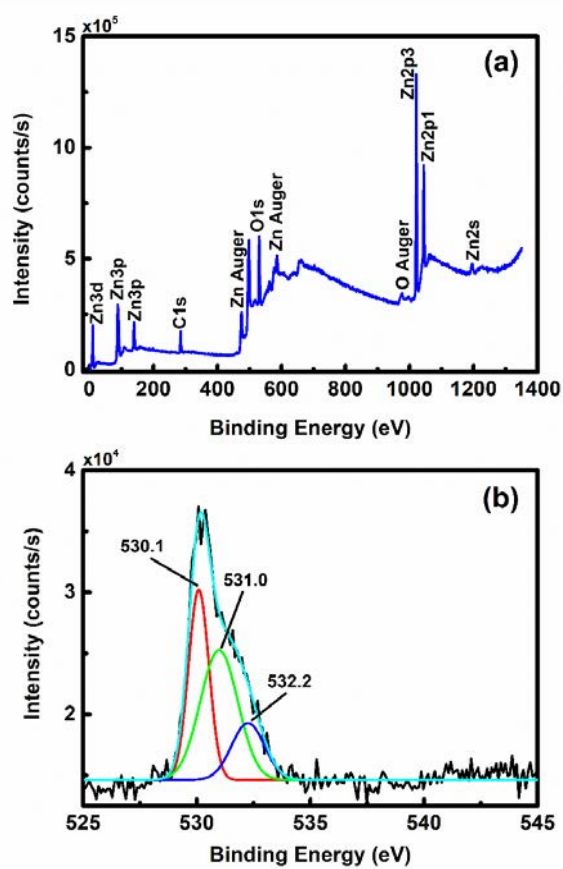

**Fig. 2.** The XPS spectra of the PED-fabricated bare ZnO thin film. A full range survey spectrum was shown in **(a)**, while high-resolution XPS chemical binding spectra O 1 s state (its deconvolution) was shown in **(b)**.

XPS analysis was performed to get information about the surface-adsorbed species and surface chemical states of the PED-fabricated ZnO thin films. Fig. 2 (a) shows XPS survey spectra, and no elements besides Zn, O and C were detected. The presence of $Zn2p_{3/2}$ and $Zn2p_{1/2}$ peaks locating at 1021.3 eV and 1044.7 eV verify that Zn exists in oxidized states. The energy difference between two Zn components is 23.4 eV, which is agreement with the reported values of ZnO [4]. Fig. 2(b) shows the high-resolution XPS spectra of O1s core level. In order to further examination, the O1s peak was deconvoluted by fitting with three Gaussian/Lorentzian peaks. The deconvoluted peaks are located at binding energies 530.1, 531.0 and 532.2 eV. Here, the lower energy peak is attributed to lattice oxygen in hexagonal wurtzite structure of ZnO [25–27]. The middle peak at binding energy of 531.0 eV is associated with oxygen vacancy in ZnO lattice which confirms the existence of surface defects on ZnO thin film. The higher energy peak corresponds to chemisorbed oxygen species such as $OH^-$, $H_2O$ and $O_2$ [4,28,29].

The PL measurements were performed using a pulsed Xenon lamp with an excitation wavelength of 350 nm at room temp. to unveil the defect states in the PED-fabricated ZnO thin films, and Fig. 3(a) shows the PL spectrum. The raw PL data were processed using baseline subtraction and 2$^{nd}$ order Savitzky-Golay smoothing, and it was deconvoluted by fitting nine Gaussian peaks to reveal hidden peaks. As seen in Fig. 3(a), the PL graph shows near-band emission at an energy of 3.17 eV (391.4 nm). Besides the near-band emission, the PED-fabricated ZnO thin films show visible emissions in the spectrum region 2.94 – 2.18 eV consisting of several distinct peaks at 2.94 eV (421.4 nm), 2.79 eV (445.2 nm), 2.71 eV (457.3 nm), 2.55 eV (486.6 nm), 2.39 eV (518.5 nm), 2.35 eV (528.4 nm), 2.29 eV (541.3 nm), and 2.18 eV (568.4 nm). Basically, transition from valance band (VB) to conduction band (CB), and VB to shallow defect levels occurs upon photo-excitation in the PL measurement, which give rise to subsequent emissions; from CB to deep levels, shallow levels to VB, shallow levels to deep levels, and hole capture at deep levels gives violet, blue and green emissions according to energy levels difference [29]. Since the excitation energy (3.57 eV) was slightly above the energy bandgap of our samples (~3.26 eV reported in our previous work [16]); therefore, the electrons excited from the VB to the CB as well as shallow defect levels.

Generally oxygen vacancies are considered as the origin of observed green emission in the PL spectrum. On the other hand, zinc related defects such as; zinc interstitials (neutral, single and double ionized) are responsible for blue emission. According to the previous reports [29–34], the observed emissions in PL spectrum (Fig. 3(a)) can be summarized as follows: the transitions from 1) excitonic states to VB, or zinc interstitial levels to VB, 2) CB to O interstitials levels, 3) double or single ionized Zn vacancy levels to VB, 4) double ionized Zn interstitials levels to VB, 5) single ionized Zn interstitials levels to neutral Zn vacancy levels, 6) single ionized O vacancy levels to VB, 7) CB to neutral antisite O levels, 8) CB to antisite

O or interstitials O levels, and 9) CB to double ionized O vacancy levels. The first four peaks (2-5) are responsible for blue emission, whereas last four peaks (6-9) are responsible for green emission. The numbers correspond to those marked peaks in Fig. 3(a).

The observed possible transitions in the PL spectrum evidence that the PED-fabricated ZnO thin films possess many defects levels within the bandgap which are induced by O and Zn vacancies, O and Zn interstitial atoms, and antisite O atoms. These defects states were unintentionally introduced the ZnO structure during PED ablation process. These defect-related states also give rise to visible-light response ability in addition to the intrinsic UV light detection capability of ZnO as reported in Refs. [14,29].

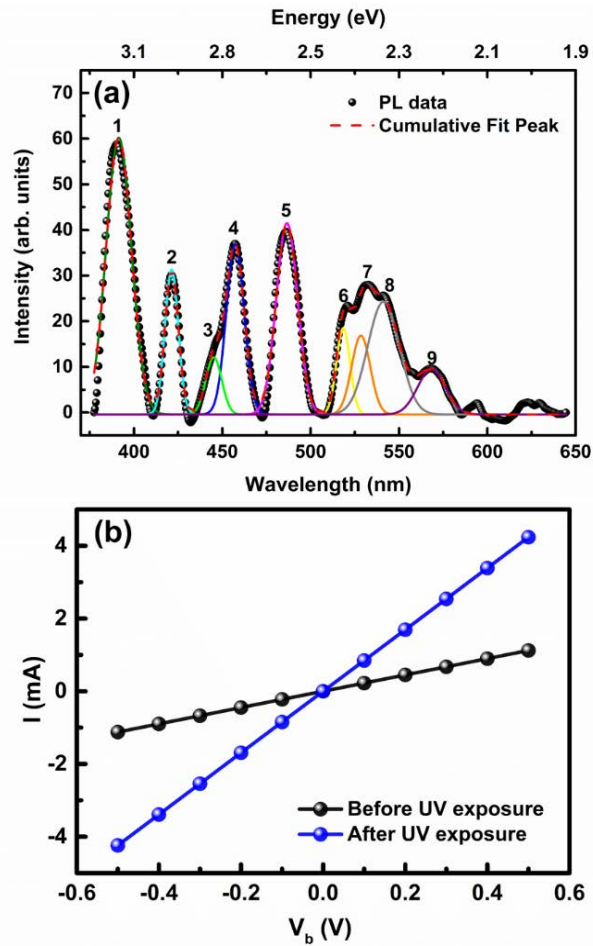

**Fig. 3. (a)** PL spectrum of B-ZnO thin film. In here, black circles correspond to the smoothed and baseline subtracted PL data. The numbers correspond to the deconvoluted peaks **(b)** I-V measurements of B-ZnO sample before and after 3 h of UV light exposure. The current data are acquired by sweeping voltage in 0.2 V steps.

Fig. 3(b) demonstrates the I-V characteristics of B-ZnO thin film device before and after UV light exposure. It was observed that the I-V curves were perfectly linear and symmetric in both bias regions (forward and reverse), indicating good ohmic contact nature.

This means that the device, which was fabricated as metal-semiconductor-metal, operates as photo-conductor. The initial resistance of ZnO thin film was determined from the slope of I-V curve as 0.45 kΩ. After 3 h of UV light illumination under high vacuum, the resistance of ZnO thin film reduced to 0.12 kΩ. The observed drop in the resistance can be explained as in the following manner. The atmospheric molecules such as $O_2$ and $H_2O$ are readily adsorbed onto surface of sample after exposure to air. These adsorbates behave like electron trapping surface states, and hence reduce the electron density of n-type ZnO material. As the UV illumination causes desorption of these adsorbates from the sample surface into vacuum, the trapped electrons are released back to the structure. Therefore, the resistance of the UV-illuminated sample becomes lower than that of as-grown ZnO thin film.

Now we turn to time-dependent photo-response measurements. All measurements were done in a high vacuum (low $10^{-3}$ Pa) to ensure a controllable environment. During these measurements, the applied voltage between source and drain was kept constant at 0.5 V. To promote desorption of adsorbates which were already adsorbed on the surface in air, the samples were exposed to UV light. The UV irradiation wavelength was specifically selected to be 254 nm (~4.9 eV) since it is energetically sufficient enough to remove chemisorbed $O_2$ and $H_2O$ molecules on the surface [19] as well as for generating electron-hole pairs in the depletion regions of ZnO thin film ($E_g$ ~3.26 eV [16]). Time-dependent current variation of B-ZnO thin film photo-conductor was shown in Fig. 4(a). The variation in the measured current $\Delta I/I_0$ (%) were determined by using the following expression;

$$(\frac{\Delta I}{I_0}) \times 100\% = (\frac{I - I_0}{I_0}) \times 100\% \tag{1}$$

where $I_0$ is the dark current read before UV exposure and $I$ is the measured current after UV exposure. Upon the UV light exposure the photo-response of sample initially increased rapidly, and a subsequent slower exponential growth was observed. After the UV light source was turned off, fast and slow exponential decay trends were observed similarly to the exponential growing parts. After 5 ks of UV illumination, the time necessary to reach a stable high vacuum condition in vacuum probe station, the variation in current of B-ZnO sample was measured as high as ~350%. This result supports the high photoconductive gain of ZnO previously mentioned in many reports [6]. After turning the UV light off, the current variation didn't decrease initial level (dark level) since the oxygen and water molecules desorbed by the UV light were pumped away, and thus, the concentration of adsorbate molecules re-adsorbed on the surface is smaller than it was prior to illumination.

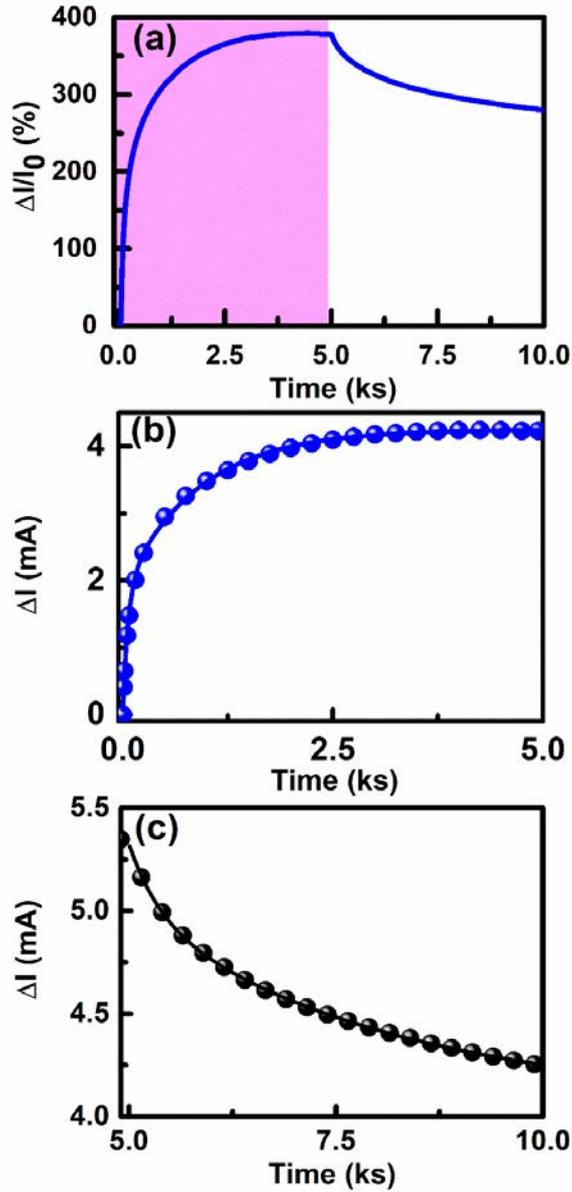

**Fig. 4. (a)** The change in the currents of B-ZnO thin film that was exposed to 254 nm UV light for a period of 5 ks under high vacuum. After a period of 5 ks, the UV light was turned off and the sample was let to relax in high vacuum for another 5 ks. The violet region shows the time interval when the UV light is on. $\Delta I = I - I_0\ (mA)$, where $I_0$ denotes the current read before UV illumination and $I$ is the current after UV illumination. During measurement, the applied voltage between source and drain was kept constant at 0.5 V. The growing part of photo-response variation of **(a)** B-ZnO thin film was shown in **(b)** with blue spheres, while the decaying part was shown in **(c)** with black spheres. Blue and red curves illustrate the fitted curves using Eq. (1) and (2), respectively.

As mentioned above, the photo-response characteristics of ZnO exhibit an exponential raise and decay behaviors determined by the UV light is turned on-off. The photo-response raised exponentially upon UV light exposure was fitted by the sum of two-exponential growth functions;

$$I = I_0 + A_1(1 - \exp(-t/\tau_f)) + A_2(1 - \exp(-t/\tau_s)) \qquad (2)$$

and it was shown in Fig. 4(b) with a blue line. In equation (2), $A_1$ and $A_2$ fit constants, $I_0$ correspond to related dark current read before the illumination, $I_0 = 4.27\ mA$ (ohmic contact leads to high dark current), and the extracted time constants were $\tau_f$=0.09 ks and $\tau_s$=0.97 ks corresponding to the first and second terms of the Eq. (2), respectively. Here $\tau_f$ is the time constant of fast component and $\tau_s$ is the time constant of slow component. When turning the UV light off, the observed decrease in photo-response was fitted by the sum of two-exponential decay functions as;

$$I = I_0 + A_1 \exp(-t/\tau_f) + A_2 \exp(-t/\tau_s) \qquad (3)$$

where $I_0 = 4.06\ mA$, and the extracted time constants are $\tau_f$=0.36 ks and $\tau_s$=3.14 ks corresponding to the first and second terms in the Eq. (3), respectively. The fitted curve was shown in Fig. 4(c) with a black line.

In our previous work [16], we have compared the photo-response properties of PED-fabricated ZnO thin film for air and high vacuum environments, and we saw that the photo-response decreases faster in air (with a ~%45 relative humidity) and reaches initial condition upon turning the light off. When the water molecules are adsorbed with the hydrogen sides on the surface, which are positively charged due to the high electronegativity of oxygen than hydrogen [15], they capture more electrons than oxygen. In addition, Li *et al.* [10] reported that water adsorption especially at high humidity cases (> 70%) leads to the fast shortening of the photocurrent in decaying stage for ZnO nanowires. Moreover, Panda *et al.* [35] reported two time constants for current decaying stage of thermally grown ZnO thin films, and they related the fast decay time to water adsorption, the slow decay rate to oxygen adsorption. According to our results and literature, we concluded that the estimated two fast time constants in above are more likely to occur with photo-assisted desorption/adsorption of $H_2O$ molecules from the surface, whereas the slow time constants are due to relatively slow desorption/adsorption rate of $O_2$ molecules from the surface.

As seen in Fig. 4(a), the growing and decaying part of photo-response variation are not symmetric. This clearly shows the effect of atmospheric adsorbates on the photo-response characteristics of our samples. In the decaying part, the fast time constant is 4 times as large relative to fast time constant for growing part (0.36 ks / 0.09 ks) corresponding to re-adsorption of $H_2O$, whereas the slow time constant is ~3.2 times higher than that of the growing part (3.14 ks /0.97 ks) as a consequence of re-adsorption of $O_2$ molecules on the surface. When the high vacuum reached its stable value (~$5.0 \times 10^{-3}$ Pa), the desorbed $H_2O$ and $O_2$ molecules from the surface with the help of UV light were mostly evacuated from the environment. Therefore, fast and slow time constants were largely increased since re-adsorption rates were greatly decreased due to lack of adsorbate molecules, and consequently an asymmetric trend was observed in the photo-response character determined by the UV light was turned on-off. This indicates that the dominant mechanism of photo-response in ZnO is adsorption/desorption of atmospheric species on the surface, and it is difficult to distinguish photo-generation/recombination times from these large time constants. Therefore, to eliminate adsorbate effects on the photo-response, we have encapsulated the surface of ZnO with a thin layer of $SiO_2$.

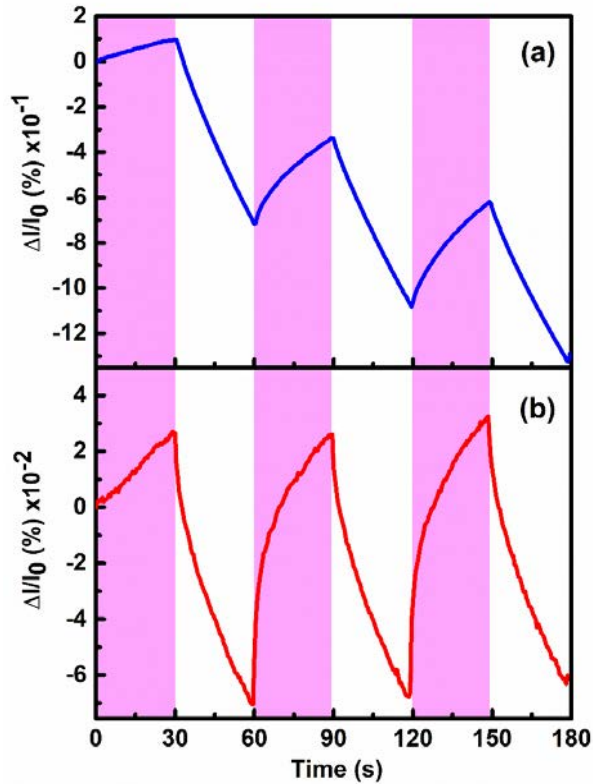

**Fig. 5.** Short-period TPS measurements of **(a)** B-ZnO thin film and **(b)** E-ZnO thin film. The violet regions show the time intervals when the UV light is on. During measurement, the applied voltage between source and drain was kept constant at 0.5 V.

In Fig. 5, we have compared the photo-response behaviors of B-ZnO and E-ZnO thin films by taking short periods TPS measurements. The TPS measurements were conducted for short periods as 30 s with three on/off cycles. Prior to the these measurements, both samples were exposed to UV light in high vacuum for duration of ~3 h to make sure that all possible adsorbates were removed greatly from surfaces. To quantitatively examine the results, short-period TPS curves were fitted using the same two-component exponential functions presented above in Eqs. (2) and (3), and thus the time constants were extracted. For the experiments, we first started to short period TPS measurements with UV light was on state, and after 30 s time interval, the shutter was closed (off state). In Fig. 5(a), following a very small but sharp drop within 3.2 s, an exponential decay trend in the measured current was observed as a consequence of the re-adsorption of a trace amount of adsorbates on the sample surfaces with an adsorption time of 74.8 s. When turning the UV light on, a very small increment occurred due to the contribution of photo-generated charge carriers within ~2.0 s, which was followed by an exponential increment in the current due to photo- desorption process with a desorption time of 35.4 s. The adsorption rate with the estimated time constant ($\tau_{ad}$= 74.8 s) was found to be smaller than desorption rate ($\tau_{des}$= 35.4 s) of B-ZnO sample. The drastic difference between the adsorption and desorption rates leads to a downward trend in the overall current variation within a couple of on/off cycles as seen in Fig. 5(a). On the other hand, after the $SiO_2$ encapsulation of ZnO surface, the sharp drop time decreased to 0.8 s as turning the UV light off, and the adsorption rate was estimated as 30.6 s. When turning the UV light on, photo-generation occurred within a 0.7 s, which was followed by an exponential increment with a desorption rate of 19.3 s, at the sensitivity limits of our measuring system. These results apparently indicate that the surface encapsulation method notably reduce the estimated time constants for both adsorption and desorption rates. Additionally, it reveals more reasonable photo-generation and recombination time constants. Besides, when the amplitudes of current variations of B-ZnO and E-ZnO samples were compared, the amplitude of E-ZnO sample was almost one order of magnitude smaller than that of bare counterpart, indicating adsorbates responsible for high photoconductivity, and the effect of adsorbate molecules were partially eliminated after encapsulation process.

The photo-response properties of ZnO thin films greatly improved after the encapsulation of surface with a thin layer of $SiO_2$. The TPS plots of E-ZnO thin film exhibit an almost symmetric trend. Once the UV light is turned off, the photo-current of the E-ZnO sample decreased to its initial level indicating enhanced stability and reversibility of our device as seen in Fig. 5(b). Despite the encapsulation significantly improve the device performance, there would be another possible mechanism that still leads to exponential behavior seen in Fig. 5(b), as explained in the discussion section.

## 4. Discussion

Possible technological applications of ZnO materials in the fields of gas sensors and optoelectronics have been investigated for many years [36], but the relation between the material surface and atmospheric adsorbates have not been fully understood yet. In fact, its photo-response is mainly governed by the charge transfer doping phenomenon that occur during adsorption/desorption of adsorbates in air as previously discussed [10,37]. When the adsorbates like $O_2$ and $H_2O$ in air are stuck on the surface of ZnO nanostructures especially with high surface area-to-volume ratio, they trap free electrons by creating a low conductivity depletion layer in the vicinity of surface. The trapped electrons are released fast at first via desorption of $H_2O$ molecules, and then continues at a slower rate due to desorption of $O_2$ molecules under UV light that promotes desorption process. As evidenced in our XPS results, the PED-fabricated ZnO thin films possess oxygen vacancies on the surface, which they serve active areas for adsorption of $O_2$ and $H_2O$ molecules. These results confirm our suggestions about adsorption/desorption mechanism in ZnO thin films with large surface area-to-volume ratios.

For ZnO thin film with inherently high surface area-to-volume ratio as a result of high energy deposition technique like PED in our case, the conductivity is mainly governed by the adsorption/desorption of atmospheric adsorbates. In the light of the experimental results obtained in 5 ks photo-response measurements (Fig. 4), the conductivity changes can be written in the form of $\Delta\sigma_{ZnO}(t) \propto \sigma_{des}(t)$ under UV illumination. Here, $\sigma_{des}(t)$ is the contribution of photo-desorption of adsorbates from surfaces to conductivity. The contribution of photo-desorption process, $\sigma_{des}(t)$, can be divided into two components as $\sigma_{des}(t) = \sigma_{H_2O}^{f}(t) + \sigma_{O_2}^{s}(t)$. Here, $\sigma_{H_2O}^{f}(t)$ is the contribution of fast component by reason of water desorption, while $\sigma_{O_2}^{s}(t)$ is the contribution of slow component by reason of oxygen desorption. These fast and slow terms vary as a function of time in the form of $\sigma_{H_2O}^{f}(t) = 1 - e^{-t/\tau_f}$ and $\sigma_{O_2}^{s}(t) = 1 - e^{-t/\tau_s}$ corresponding to the first and second terms in the right side of Eq. (2), respectively. On the other hand, upon turning the UV light off the alteration in conductivity occurred as $\Delta\sigma_{ZnO}(t) \propto -\sigma_{ad}(t)$. Here $\sigma_{ad}(t)$ is the contribution of the re-adsorption of adsorbates on the surface of film. The same two components (fast and slow) aforementioned are responsible for exponential decay of photo-response. Therefore, fast and slow decay processes change as a function of time $\sigma_{H_2O}^{f}(t) = e^{-t/\tau_f}$ and $\sigma_{O_2}^{s}(t) = e^{-t/\tau_s}$, respectively.

To get deeper insights into the photo-response character of ZnO thin film under UV light, we have conducted short period TPS measurements. The results reveal that the conductivity variations can be written in the form of $\Delta\sigma_{B-ZnO}(t) \propto \sigma_{ph} + \sigma_{des}(t)$ under UV exposure, $\sigma_{ph}$ is the contribution of photo-induced charge carriers in the depletion layers of ZnO thin film. When the UV light is turned off, the conductivity changes can be written as $\Delta\sigma_{B-ZnO}(t) \propto -(\sigma_{ph} + \sigma_{des}(t))$. The observed exponential trend proves the $\sigma_{ph} \ll \sigma_{des}(t)$ due to surface layers that absorb most of UV photons for desorption process.

In this work, we showed an experimental method that inhibits the interactions between atmosphere and surface simply by a $SiO_2$ encapsulation layer. Therefore, we assumed that the contribution of adsorbates to the photo-response character was completely blocked by means of encapsulation. However, despite the effect of adsorbates on the photocurrent variation was inhibited, the exponential trend was still observed in Fig. 5(b). As known in literature, the existence of defects in the crystal structure would introduce several defect-induced states acting as hole trap centers within the bandgap of any semiconductor [14,38,39]. Also, oxygen vacancies are the predominant type of defect in metal oxides semiconductors such as $In_2O_3$ and ZnO thin films especially produced by PLD or PED techniques [20,40]. Our PL and XPS results confirm the existence of many defective states both on the surface and within the bandgap of PED-fabricated ZnO thin films. Therefore, even if we isolate the surface of ZnO thin film from the environment, the photo-generated holes are mostly trapped at these defect states leaving behind unpaired electrons in the ZnO that still contribute to the photocurrent. Due to the lifetime of these unpaired electrons is further increased by trapped holes; the recombination rate of electron-hole pairs is progressively slowing down. The additional contribution of trap-states to the conductivity ($\sigma_{trap}$) plays another role on the overall conductivity variation of ZnO thin film ($\Delta\sigma \propto \sigma_{ph} + \sigma_{trap}$) determined by the UV light is on-off. Eventually, ZnO thin film shows slow photo-response to UV light leading to exponential behavior as still seen in Fig. 5(b).

## 5. Conclusion

In summary, we investigated the mechanism behind slow photo-response behavior of the PED-fabricated ZnO thin films. The photo-response of ZnO occurs because of two mechanisms acting simultaneously, which make it difficult to pick out the individual contributions since the effect of adsorbates is dominant. To eliminate the effect of adsorbate, we encapsulated the surface of ZnO with a thin layer of $SiO_2$. We see that defect-induced hole trap-states plays another role on the overall conductivity variation of ZnO thin film, which still leads to slow photo-response to continue. Therefore, controlling intrinsic defects during material growth process as well as encapsulation process is important to obtain fast speed UV photodetector fabricated out of ZnO material.

We believe that our results lead the way for future application of high speed and sensitive ZnO photodetector with choosing effective encapsulation materials and prompt for further investigation in encapsulation of other n-type materials.

## Acknowledgments

The authors would like to thank Serap Yiğen for the encapsulation experiments, and S. Batuhan Kalkan for the experimental setup. Characterizations of samples were conducted in the Center for Materials Research of İzmir Institute of Technology (IYTE-MAM) and Ege University Central Research Test & Analysis Laboratory Application & Research Center (EGE-MATAL). A part of this work has been supported by Ege University Scientific Research Project (BAP) with Project No. 15-FEN-058.